\documentstyle[epsf]{aipproc}
\newcommand\movit{\par\hangindent=33truept\hangafter=1}
\newcommand\hten{\hskip10truept}

\begin{document}

\title{LUMINOSITIES AND STAR FORMATION RATES OF GALAXIES OBSERVED WITH
THE ULTRAVIOLET IMAGING TELESCOPE: \\
A Comparison of Far-UV, H$\alpha$, and
Far-IR Diagnostics}

\author{Michael N. Fanelli $^{* \dagger}$, Theodore P. Stecher $^*$, \\
and the UIT Science Team}

\address{$^*$ Laboratory for Astronomy and Solar Physics\\ NASA-Goddard Space
Flight Center \\  Greenbelt, Maryland 20771\\
$^{\dagger}$ Hughes STX Corp, Lanham, MD 20706}

\maketitle

\begin{abstract}
During the UIT/Astro Spacelab missions, the {\it Ultraviolet
Imaging Telescope} obtained spatially resolved 
far-UV ($\lambda\lambda\ 1500$~\AA) imagery
of $\sim 35$ galaxies exhibiting recent massive star formation.  The sample
includes disk systems, irregular, dwarf, and blue compact galaxies.
The objects span an observed FUV luminosity range from $-17$ to $-22$ 
magnitudes. We estimate global star formation rates by comparing the observed
FUV fluxes to the predictions of stellar
population models, and compare the FUV-derived astration rates to those
derived from H$\alpha$ and far-IR photometry. 
\end{abstract}

\section*{Motivation}

Many aspects of galaxy evolution are driven by the
spatial distribution and formation history of high-mass (M $ > 3 $~M$_\odot$) 
stellar populations.  Massive OB stars emit most of the radiation in the
vacumn ultraviolet ($\lambda < 2500$~\AA) while cooler, solar-type stars
emit minimal radiation at these wavelengths.
Therefore UV imagery provides a snapshot of the recently formed
stellar populations in galaxies.  Prior to 1990, only limited samples of UV 
photometry were available for galaxies, consisting mostly of integrated fluxes
\cite{code82,donas87,buat92} and modest resolution images derived from
sounding rockets, e.g., \cite{jhill84}.

\section*{The Data}

During the Astro/UIT Spacelab missions in December, 1990 and March, 1995
 the {\it Ultraviolet
Imaging Telescope} \cite{stecher92} obtained deep, high spatial 
resolution ($\sim$3$^{\prime\prime}$, FUV ($\lambda\lambda\ 1500$~\AA) imagery
of $\sim 35$ galaxies exhibiting recent massive star formation.  
 The images cover the full angular extent of each
system, most of which have diameters exceeding  $ 5^{\prime}$.  The spatial
resolution of
the FUV images is an improvement of $\approx 5$--20 over previously available
data.    These data permit determination of both global FUV
properties with improved photometric precision, and detailed
investigation of galaxian morphology at intermediate (spiral arms, nuclear
rings) \cite{waller97,rwomar96,dsmith96,fanelli97}, 
and small (star-forming complexes) \cite{sneff97} scales.

\section*{Results}

We derive global star formation rates by comparing the observed
FUV fluxes to the predictions of stellar
population models. For this interation, a simple model was chosen:
a power law IMF with slope = $-$1.35, solar abundances, and mass range,
$1 < $~M\,/\,M$_\odot < 100$.   Assuming continuous star formation, the
observed flux  can be compared
to the model FUV luminosity  to derive a star formation rate.  
Astration rates derived from FUV data can be compared to those
derived from H$\alpha$ and far-IR fluxes to explore the utility of
star formation rates estimated from FUV data, and the star formation
history of these systems (Table~1).

\begin{table}[hb]
\caption{Star Formation Rates (M$_\odot$ yr$^{-1}$)}
\label{mnf:table1}
\begin{tabular}{lddd}
Rate & NGC 3310 & NGC 4214 & NGC 4038/9 \\
\tableline
 $\dot {\rm{M}}$(FUV) & 1.3 & 0.20 & 1.6 \\
$\dot {\text{M}}$(H$\alpha$) & 3.7 & 0.39 & 6.6 \\
$\dot {\text{M}}$(FIR) & 8.6 & 0.50 & 14.4  \\
\end{tabular}
\end{table}

\movit{$\bullet$ \hten The objects span an \underbar{observed} (uncorrected 
for internal extinction) FUV luminosity
range from $-17$ to $-22$ magnitudes (Figure~1). }
 
\movit{$\bullet$ \hten The global (FUV--V) colors span a range from $-1.3$ to
 $+3.6$, a much larger range than that found using optical bandpasses
 alone. }
 
\movit{$\bullet$ \hten Late Hubble types are bluer in these colors,
as expected.}
 
\movit{$\bullet$ \hten Massive star formation rates of $0.03 \lesssim $ SFR
$\lesssim 4 $ M$_\odot$~yr$^{-1}$ are found based on the 
\underbar{observed}  FUV luminosity.}
 
\movit{$\bullet$ \hten For the Sm/Im galaxy NGC~4214 the ratio of
FIR\,/\,FUV star formation rates is $\sim 2.5$ indicating that the
 FUV emission directly traces a significant fraction of the recently formed
 high-mass stars.}

\movit{$\bullet$ \hten For the bluest systems, the ratio of 
FUV\,/\,H$\alpha$ astration rates is found to be comparable.}
 
\movit{$\bullet$ \hten In some dusty, FIR-luminous systems, substantial FUV
light is observed, e.g., the merging system NGC 4038/39 (the ``Antennae'') 
\cite{sneff97}.  Although the FIR\,/\,FUV astration rate ratio is $\sim 10$,
the detection of extensive FUV emission indicates that massive star 
formation can be \underbar{\bf directly} 
probed in these systems, despite the presence of significant extinction.}

\vskip10truept

\begin{figure}[tbd]
\centerline{\epsffile{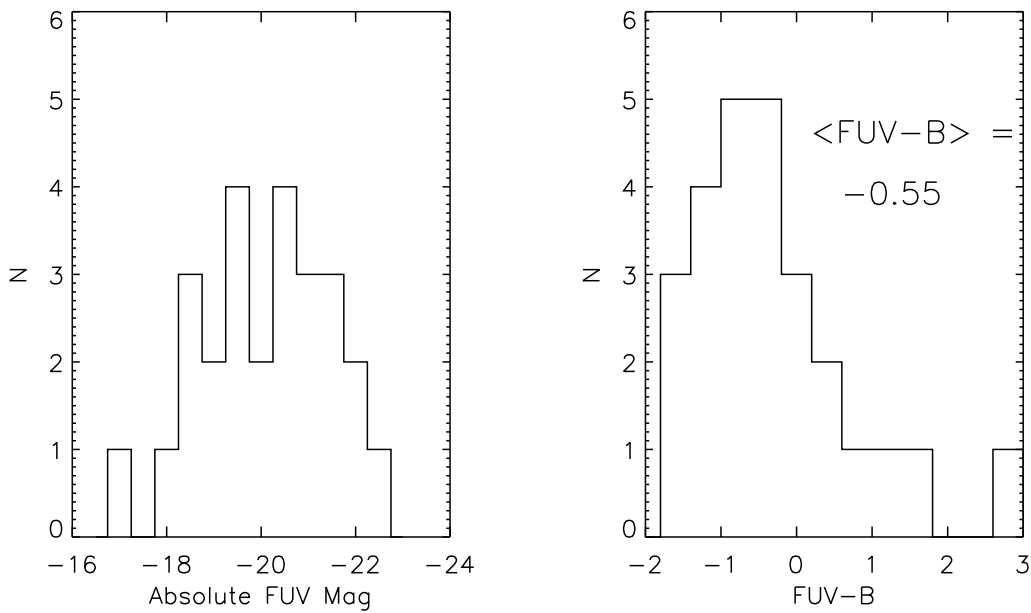}}
\vspace{15pt}
\caption{The distribution of absolute FUV magnitudes (left) and FUV to optical
band colors (right) for the sample of disk galaxies observed by UIT.}
\end{figure}

\section*{The UIT Galaxy Atlas}

We \cite{marcum97}
 are constructing an Atlas which combines the FUV imagery obtained
by the {\it Ultraviolet Imaging Telescope}\/ and associated optical imagery
obtained at ground-based telescopes. Our primary goal is to provide a
morphological Atlas of Galaxies extending from far-UV
($\lambda\lambda \sim$\,1500~\AA) to near-IR
($\lambda\lambda \sim$\,2.2$\mu$) wavelengths.  Comparison of the UV images
 with images at visible/NIR wavelengths will provide
 critical information on the intensity, spatial pattern, and temporal evolution
 of the massive stellar populations;  the distribution  of dust along the 
arms and central bars of spiral galaxies, and the relationship
 between recent star formation and the interstellar medium.

\begin{figure}[hb]
\centerline{\epsffile{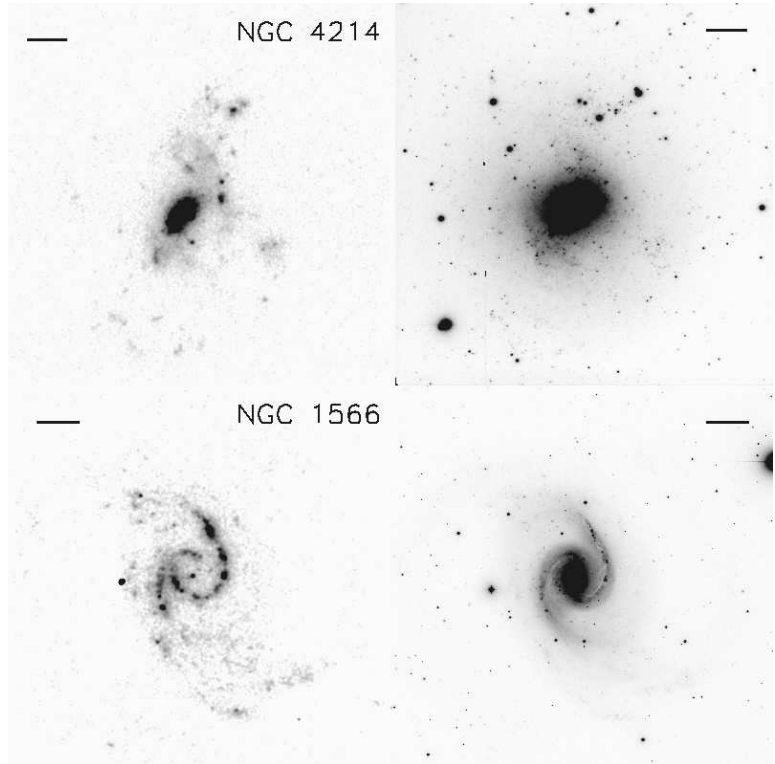}}
\vspace{20pt}
\caption{A mosaic of far-UV and I-band imagery of two disk galaxies observed
by the {\it Ultraviolet Imaging Telescope}. FUV ($\lambda\lambda \sim 1500$ 
\AA) images from UIT are displayed on the left, ground-based I-band images on
the right. All images are displayed in a north up, east left orientation.
The horizontal bar indicates an angular scale of
60$^{\prime\prime}$, corresponding to a physical scale of 1.2 kpc 
for NGC~4214, and 5.1 kpc for NGC~1566.}
\end{figure}

\end{document}